# Heating of the interstellar gas by cosmic rays and warm transparent ionized plasma observed by pulsar dispersions

Y. Ben-Aryeh

Physics Department, Technion-Israel Institute of Technology, Haifa, 32000. Israel

E-mail: phr65yb@physics.technion.ac.il   ,   ORCHID: 000-0002-6702-2530

ABSTRACT

Electron densities in different locations of our galaxy are obtained in pulsar astronomy by dividing the Dispersion Measure (DM) by the distance of the pulsar to Earth. The properties of the interstellar plasma are related to its heating by cosmic rays. Following the present analysis DM's measurements are obtained with different properties for different temperatures in three regions: 1) For relatively low temperatures the state of molecular, atomic, and ionized Hydrogen was analyzed by the "interstellar medium" (ISM) model with partially ionized plasma. In this region various spectroscopic effects are obtained. 2) For temperatures approximately above $20000\,(K)$ the interstellar gas was found to be completely ionized medium and this plasma is defined as warm ionized plasma (WIM) where this plasma is transparent. This property is obtained from solution of Saha's equation in which the index of refraction is real, but the plasma can be observed by dispersion measurements. For very high temperatures defined as hot ionized matter (HIM), X-rays are obtained by the plasma. We concentrate in the present work on the analysis of WIM plasma. We calculate the mass densities of such plasma and compare it with dark matter mass densities which are found to be larger. But some factors which may reduce this difference are discussed.

## 1. Introduction

Our analysis is based on the idea that the heating of the plasma is obtained by cosmic rays interacting with these plasmas [1-3]. For electrons densities above the critical value $n_{e,crit} = 0.1\,(cm^{-3})$ the cosmic energy is going for ionizing the gas and in this region partially ionized plasma is produced with corresponding spectroscopic measurements as analyzed by the ISM model [4]. For small electrons densities ($n_e < 0.075 (cm^{-3})$) most of the cosmic radiation is going for heating the ionized gas while only small fraction of this energy is given for ionizing the low density stellar gas and this fraction becomes smaller as the density $n_e (cm^{-3})$ becomes smaller. We referred to this region as the transparent warm ionized medium (WIM) but at very high temperatures (around $10^6 - 10^7 (K)$) X-rays are emitted and we referred to such plasmas as hot ionized medium (HIM). Following this idea, we find that the temperatures of the interstellar gas are functions of their densities, and such functions are different for the above three regions of the interstellar plasma analyzed as follows.

In the range of temperatures 10-10000 (K) the state of molecular, atomic, and ionized Hydrogen was analyzed by the "interstellar medium" (ISM) model for corresponding densities in the range $0.2 - 10^6 \left( cm^{-3} \right)$ in which the interstellar gas is partly ionized [4]. For example, for temperatures $8000 - 10000(K)$ we have approximately ionization rate of 10% and in this range of temperatures one gets a lot of spectroscopic data. By considering the heating of interstellar medium by cosmic rays the temperature in this ISM region was found to be proportional to $(N/n)^2$ where $N$ is the density of the cosmic rays and $n$ is the density of the interstellar gas [1,2]. The proportionality constant was estimated to be given as: $2 \cdot 10^{14}$ [1].

According to the present analysis there is a phase transition at temperature around $20000(K)$ from partly ionized Hydrogen plasma (which is the main component of the stellar gas) to completely ionized plasma. We find also that the completely ionized low-density plasma is transparent so that spectroscopic data are not available in this region (besides dispersion effects in the far radio region). Quite long ago the possibility of having interstellar galactic "Corona" was studied [3]. The density of ionized electrons and protons in the Corona was obtained as $n_e = n_p = \dfrac{C}{T} (cm^{-3})$ where the constant was estimated as C=500. The extension of the corona perpendicular to the galactic plane was studied. Certain doubts on the possibility to measure the Corona were raised by Spitzer [3]: "It is evident that the intensity of the Coronal emission, when reduced by so large a factor, is entirely ununobservable". He also discussed the possibility to measure the Corona by absorption lines of heavier atoms but commented: "it is uncertain whether such radiation shortward of the Lyman limit can reach Earth in view of probably heavy absorption by neighboring clouds".

In recent article [5] extensive calculations have been developed to mediate the ways by which cosmic rays affect the evolution of the interstellar medium. In Figure 1 of this article, it has been shown that density of the electrons (in unit of $g\,cm^{-3}$) is an inverse linear function of temperature (with some fluctuations around this linear relation). By changing the unit of density to $cm^{-3}$ we get the approximate relation from this table as

$$n_e = \frac{1500}{T}(cm^{-3}) \quad ; \quad T = 20000\,(K) \rightarrow n_e = 0.075\,(cm^{-3}) \tag{1}$$

where the constant 1500 is larger by factor 3 relative to that estimated by [3] and there are fluctuations around this linear relation.

Eq. (1) is in fair agreement with our analysis for the WIM and HIM mediums. The transition around the temperature $20000\,(K)$ from the partially ionized Hydrogen plasma to completely ionized plasma corresponds to average electrons density value of $0.075\,(cm^{-3})$. The HIM Plasma at temperature $10^6 - 10^7\,(K)$ corresponds to electrons density of $0.0015 - 0.00015\,(cm^{-3})$, respectively. We find that in WIM and HIM regions the electrons density is inversely proportional to its temperature. This result follows from the condition that most of the energy transferred from the cosmic rays leads to heating of the ionized gas (a very small fraction of this energy is sufficient for ionizing the low-density gas).

In the present work we would like to relate the properties of the warm ionized mater (WIM) to pulsar dispersion measurements. Pulsars are usually produced by rotating neutron stars emitting beams of electromagnetic radiation with radio frequencies which sweep around in space as the neutron star rotates. Detection of such regular pulses of radiation can be obtained when any of such beams crosses the line of sight to the Earth. An electromagnetic wave passing through ionized plasma includes a frequency dependent change in the group velocity, a phenomenon known as dispersion. This induces an additional time delay $t_d$ for the pulse time arriving on Earth which can be approximated in cgs units as [6-9]:

$$t_d = \frac{e^2}{2\pi m_e c} \frac{DM}{v^2} \qquad (2)$$

where $e$ and $m$ are the electron charge and mass, respectively, $c$ velocity of light, $v$ the frequency, and DM is the "dispersion measure" defined as:

$$DM = \int_0^d n_e dl . \qquad (3)$$

Here $n_e (cm^{-3})$ is the electron density and $d$ is the distance from the neutron star to Earth. The net result is that when astronomers observe radio pulsars the pulse is delayed, and the amount of delay depends on the radio frequency and DM. The time delays of such pulsars are obtained due to long astronomical trajectories over low density plasma which leads to broadening of such pulses and time delays as function of frequency (i. e. dependence of the index of refraction on frequency). The time delays in pulsars are proportional to the product of average ionized electrons density and the length of the trajectory of the pulse before arriving on Earth. Extensive literature on measurements of DM by pulsars was published and there are many discussions on various effects related to such measurements [6-9].

In the present work we obtain average electrons densities ($n_e / cm^3$) and correspondingly average mass densities ($kg/m^3$) in different locations in our galaxy, the Milky Way, by using results from pulsars astronomy. There are many measurements of DM but for getting average densities we need to divide them by the distance of the pulsar to Earth and accurate distances to astronomical objects are difficult to obtain.

We treat in the present work tenuous transparent warm stellar gas where its temperature is above $20000(K)$ so that Hydrogen is ionized completely but below $10^6(K)$. The lower boundary is obtained by the requirement that the plasma is completely ionized, and the upper boundary is fixed by the requirement that bremsstrahlung effects can be neglected. For this warm ionized medium (WIM) the index of refraction is real and the plasma is transparent but can be measured by the dispersion measure (DM) defined by Eqs. (2-3). Also, this model is valid for electron densities which are smaller from $0.1\,(cm^{-3})$

Quite long ago Van Vleck [10] explained by following Bohr-Van-Leeuwen theorem [11,12] that when classical mechanics including Boltzmann statistics is valid the averaged value of the magnetic field vanishes. Such local classical kinetic equilibrium (LCKE) is valid under the conditions: 1) quantum and relativistic effects are negligible in the plasma. 2) None of the distribution functions are time dependent. 3) The distribution functions are Maxwellian. Such conditions are valid along the extreme path length in the region of warm ionized transparent medium (WIM) with low electrons densities. While the present free electrons model is based on the 3 conditions explained above, quantitative results for this model can be obtained from the book of Rybicki & Lightman [13] where in this book this model is termed as "Thermal Free-Free Emission". The relevant equation is derived by integrating the bremsstrahlung emission over the whole spectrum given as

$$J_{br} \approx 2.4 \times 10^{-27} T^{1/2} n_e n_i Z^2 \quad (erg\,s^{-1}cm^{-3}) \,. \tag{4}$$

Here $n_e$ and $n_i$ are the electrons and ions densities, respectively, $T$ is the absolute temperature and $Z$ is the atomic number ( Z=1 for Hydrogen) . For simplicity all constants in this equation were converted into a number and quantum correction $g_{ff}$ which is of order 1 is omitted. We find here that the bremsstrahlung radiation emission is proportional to the density of electrons squared and to $T^{1/2}$ so it is quite simple to put numbers and find that for very low densities and warm temperature (in the WIM region) bremsstrahlung effects can be neglected. For high temperatures for which relativistic effects become important this model breaks down, and emission of X-rays by this medium defined as HIM becomes significant. The existence of such plasma was explained by the cosmic rays which heat the low-density plasma to very high temperatures.

The average electrons density obtained by 37 pulsars [14] is given as $n_e(cm^{-3}) = 0.015$. This result was obtained by methods which are independent of the electrons density model (e. g. by parallax measurements) and as given in [14], these results are approximately equal to those obtained by other authors. But electrons density measurements of 41 pulsars [15] give an approximate average value of $n_e(cm^{-3}) = 0.03$ which is two times larger. While models for warm ionized gas were developed [16] the present analysis is different as it is based on a region of warm low-density plasma that does not show absorption or emission processes (besides dispersion effects in the far radio region).

The present model and that of the ISM model are related to the properties of the dielectric constant $\varepsilon(\omega)$. i.e. *to the polarization properties of the electric field.* The present model about low density transparent ionized plasma (WIM) is based on conventional methods (see Refs. [17,18]) by which the dielectric constant is real, and the DMs measurements are related only to small changes in the index of refraction $\eta$. We should notice that the ISM model treats high electrons density for partly ionized plasma where the dielectric constant $\varepsilon(\omega)$ is mainly imaginary and it leads to absorption and emission where e. g. it can be detected through the H-alpha line.

The optical properties of the low-density transparent plasma can be related to the use of Saha's equations [17,18]. These equations were used for calculating ionization degree in thermal plasmas, and for Hydrogen plasma which is the main component of stellar plasma these equations are amenable to analytical solutions, as there is only one Saha's equation to solve. I give an analysis for the invisibility of the present plasma in Section 2. I show in this Section that the Hydrogen plasma with density below $n_e = 0.1(cm^{-3})$ is completely ionized at warm temperatures (i. e. above approximately $20000\,(K)$) The main idea following from our analysis is that the present free electrons region is involved with corresponding nucleons mass density although not observed by EM interactions (besides dispersion in the radio frequency region ).

The paper is arranged as follows: In Section 2 we treat the optical properties of the plasma in the WIM region as function of its density and temperature. In Section 3, we analyze the nucleon mass densities obtained from pulsars astronomy in the WIM region and compare the results with dark matter densities. In Section 4 we summarize our results and conclusions.

## 2. Ionization and transparency of Hydrogen plasma as function of its density and temperature

For warm completely ionized low-density plasma, under the conditions of classical free electrons model its absorption and emission are negligible. In the present Section we analyze the conditions for the transparency of such plasma. The Boltzmann distribution can be applied to describe ionization equilibrium $A^+ + e \Leftrightarrow A$ [17,18]:

$$\frac{n_e n_i}{n_a} = \frac{g_e g_i}{g_a}\left(\frac{2\pi m_e k_B T}{h^2}\right)\exp\left(\frac{-E_I}{k_B T}\right) \ . \tag{5}$$

Here $A$ is a neutral atom, $A^+$ is this atom with one ionized electron, with charge $e$, $E_I$ is the ionization potential, $k_B$ is the Boltzmann constant, T is the temperature and $g_a, g_i$ and $g_e$ are the statistical weights of atoms, ions and electrons; $n_a, n_i$ and $n_e$ are their number densities and $m_e$ is the electron mass. This equation is known as Saha's equation and is widely used for calculating ionization degree in thermal plasmas.

The simplest case of pure Hydrogen plasma is amenable to analytic solution. For this case the number of ions is equal to the number of electrons i. e., $n_e = n_I$. For Hydrogen plasma Eq. (5) is reduced to:

$$\frac{n_i}{n_a} = \frac{1}{n_i}\left(\frac{2\pi m_e k_B T}{h^2}\right)^{3/2} \exp\left(\frac{-E_i}{k_B T}\right). \tag{6}$$

Substituting numerical values, we get [17]:

$$\frac{n_i}{n_a} = 2.4 \cdot 10^{21} \frac{T^{3/2}}{n_i} \exp\left(\frac{-13.6 eV}{k_B T}\right). \tag{7}$$

where $\frac{n_i}{n_a}$ is the ratio between the number of ionized electrons $n_i(m^{-3})$ and the number of neutral hydrogen atoms $n_a(m^{-3})$. For temperature of 20000(K) and ionized electron density $n_i = 1 (cm^{-3})$ (one ionized electron per $cm^3$) we have approximately $6 \times 10^{18}$ ions per one neutral hydrogen atom!, and for higher temperatures and lower densities this ratio is increased much further. Under this condition absorption and emission of radiation related to spectral lines can be neglected. One should take into account that although in this calculation the thermal energy might be small relative to the ionization energy we get yet complete ionization due to the effective statistical weight of the continuum spectrum which is very high (proportional to $\left(\frac{2\pi m_e k_B T}{h^2}\right)^{3/2} = \left(\frac{1}{\lambda_{DB}}\right)^3$ where $m_e$ is the electron-mass, $h$ is Planck constant, $\lambda_{DB}$ is the De Broglie wavelength and the proportionality constant in Eq. (6) is obtained by using this density of states. One should consider that at the same temperature for which low density plasma is transparent the plasmas which have higher electron densities can have strong absorption and emission of radiation. This fact is related to the $1/n_i$ dependence in Eq. (6) where for plasmas with high density $n_i$ becomes very large reducing very much the ratio between the density of ionized electrons and the neutral atoms. Such plasmas might therefore lead at the same above range of temperatures to emission and absorption of radiation related to spectral lines of Hydrogen and other atoms included in the high-density stellar atmospheres.

Low density plasma is described in the present work as a plasma for which the density of ionized electrons is smaller than 0.1 ionized electron per $cm^3$. I have shown above by using Saha's equation that for temperatures above 20000 (K) the amount of neutral Hydrogen atoms is completely negligible. Then the Hydrogen plasma is composed of protons and electrons (without any neutral Hydrogen atoms). We assumed in the above discussion a stationary state for which free electron model with Boltzmann distribution is valid. Although we presented the calculations for ionization of Hydrogen plasma, which is the main component of the low-density stellar plasma we expect that plasmas which includes mixtures of Hydrogen with other atoms will follow similar results.

## 3. Low electrons densities in pulsars trajectories with nucleon mass densities compared to dark matter densities

For the low-density plasma in which the pulsars are transmitted the dielectric constant $\varepsilon(\omega)$ is real, and the index of refraction $\eta(\omega)$ is given by its square root so that we get [17,18]:

$$\varepsilon(\omega) = \left(1 - \frac{\omega_p^2}{\omega^2}\right) \quad ; \quad \eta(\omega) = \sqrt{\left(1 - \frac{\omega_p^2}{\omega^2}\right)}. \tag{8}$$

Here, $\varepsilon$ is the real dielectric constant, $\omega \, (rad/\sec)$ is the EM frequency, $\omega_p^2$ is the plasma frequency squared and the subscript $p$ refers to plasma. The plasma frequency can be obtained by using the following relation (in cgs units):

$$\omega_p^2 = \frac{4\pi n_i e^2}{m_e} \cong 3.18 \cdot 10^9 n_i \quad (cgs \; unit). \tag{9}$$

Eq. (9) leads to changes of the EM dielectric real constant of the plasma (changes in the index of refraction) above and near the plasma frequency. Below the plasma frequency there is no transmissions (only evanescent waves can exist). Under the conditions of Eq. (8) and the approximation $\frac{\omega_p^2}{\omega^2} \ll 1$ we get for the index of refraction:

$$\eta = \sqrt{1 - \frac{\omega_p^2}{\omega^2}} \cong 1 - \frac{1}{2}\frac{\omega_p^2}{\omega^2} \qquad (10)$$

The plasma frequency, $\omega_p$, is related to the density of ionized electrons as given by Eq. (9). For very low ionized electrons density this frequency is moved into the far radio frequency region satisfying the relation $\frac{\omega_p^2}{\omega^2} \ll 1$ with index of refraction which is a little below 1. Within the above approximations the phase velocity $v_p$ and the group velocity $v_g$ are given, respectively, by: $v_p = \frac{c}{\eta} \approx c / \left(1 - \frac{1}{2}\frac{\omega^2}{\omega_p^2}\right)$; $v_g \approx c / \left(1 + \frac{1}{2}\frac{\omega^2}{\omega_p^2}\right)$. The effects of small changes in the index of refraction depend on the length of EM wave trajectory so for astronomical long distances even very small changes in the index of refraction are observed in pulsars astronomy.

The pulse travel time $t_p$, and the pulse time delay $t_d$ are functions of the distance $d$ from the observer to the point where the pulsar is created, and they can be related to the group velocity as

$$t_p \approx \frac{d}{c}\left(1 + \frac{1}{2}\frac{\bar{\omega}_p^2}{\omega^2}\right) \quad ; \quad t_d = \frac{d}{c}\frac{1}{2}\frac{\bar{\omega}_p^2}{\omega^2}, \qquad (11)$$

where $\bar{\omega}_p$ is the average plasma frequency along the pulse trajectory which can be calculated according to Eq. (9) as

$$\bar{\omega}_p^2 \equiv \frac{4\pi}{d}\int_0^d \frac{n_i(l)e^2}{m_e} dl \qquad (12)$$

where $n_i(l)\ (cm^{-3})$ is the density of ionized electrons as function of the distance $l$ along the pulse trajectory. By substituting Eq. (12) into the time delay of Eq. (11) and changing the frequency to $\nu = \frac{\omega}{2\pi}$ we get Eq. (2) for pulsars astronomy. The dispersion measure $DM = \int_0^d n_i(l)dl$ is the integrated density of ionized electrons between an observer and the pulsar creation point. In pulsars astronomy the units in Eq. (2) are changed so that the unit of length for DM is given by $pc \approx 3.09 \times 10^{16}\ m(meter)$ and the unit of frequency is given by $1\ MHZ$. Then Eq. (2) can be written as:

$$t_d = 4100 \left(\frac{DM}{cm^{-3} pc}\right)\left(\frac{v}{1\ MHz}\right)^{-2}\ (sec)\ . \tag{13}$$

The dispersion measure DM is calculated by using the difference in time delay between two frequencies, in the same pulsar bandwidth:

$$t_{d,2} - t_{d,1} = 4100 \left(\frac{DM}{cm^{-3} pc}\right)\left[\left(\frac{v_1}{1\ MHz}\right)^{-2} - \left(\frac{v_2}{1\ MHz}\right)^{-2}\right]. \tag{14}$$

The dispersion measure (DM) is then given by:

$$\left(\frac{DM}{cm^{-3} pc}\right) = (t_{d,2} - t_{d,1}) / \left\{4100\left[\left(\frac{v_1}{1\ MHz}\right)^{-2} - \left(\frac{v_2}{1\ MHz}\right)^{-2}\right]\right\}\ . \tag{15}$$

There is a lot of literature on dispersion measure (DM) in various fields of pulsars astronomy. Our interest in the present work is in the average nucleons mass densities which can be obtained in different locations in our galaxy from pulsar astronomy.

In a previous paper [19] I have shown that dark Halo might be produced by low density plasma. It is of interest to estimate the average mass density of the present low-density plasma and compare it with dark matter densities in our galaxy. The dark matter is observed by gravitational effects produced by such matter which does not include electromagnetic interactions. The physical composition of this dark matter is not known and there is much speculation on the physical nature of this matter. Reference [20], for example, includes interesting discussions about dark matter. We find by the present analysis that the low-density plasma obtained by DM includes a large region of transparent plasma that might be considered as dark matter region. So, that we compare the mass densities of low-density plasma with the dark matter mass densities as given by Table 1.

| PLACE | Density GeV$cm^{-3}$ | Density $M(sun)pc^{-3}$ | Density $kg/m^3$ |
|---|---|---|---|
| LAMOST DR5& Gaia DR2 [21] | 0.532 | 0.0133 | 9.0174 x$10^{-22}$ |
| DM halo [18] | 0.2-0.4 | 0.005-0.01 | 3.39-6.78 x$10^{-22}$ |
| Sun's Location [23] | 0.43 | 0.0114 | 7.70x$10^{-2}$ |
| K-dwarfs Gaia DR2 [24] | 0.439 | 0.0116 | 7.86x$10^{-22}$ |

**Caption to Table 1**

In the first column of this table the location of the dark matter measurement is given with corresponding Refs. [21-24]. In columns 2 and 3 the dark matter densities are given in units $Gev\,cm^{-3}$ and $M(sun\,kg)\,pc^{-3}$, respectively. In the last column of this table the dark matter densities are given with unit $kg/m^3$.

The average electrons density obtained by 37 pulses trajectories [14] is given as $n_e(cm^{-3}) = 0.015$. The average mass density of these plasmas is given for Hydrogen plasma by $0.015 \times 10^6 \times 1.67 \cdot 10^{-27} = 2.505 \times 10^{-23} \, (kg/m^3)$. This analysis was made for Hydrogen plasma which is the main component of the stellar plasma. But the mass component of the real mass densities of the low-density plasma, is larger by approximate factor 1.5 relative to that calculated for Hydrogen plasma so that the mass density becomes $3.758 \times 10^{-23} \, (kg/m^3)$. We compare this value with some values of dark mater mass densities given in Table 1 by using Refs. [21-24]. We find by this table an average dark matter mass density around $7.5 \times 10^{-22} \, (kg/m^3)$ which is larger approximately by factor 20 relative to the low-density plasma mass.

The average electrons density obtained by 41 pulses trajectories [15] is given as $n_e(cm^{-3}) = 0.03$ which is larger by approximate factor 2 relative to the measurements made in [14]. So, that the average mass density for this low-density plasma is smaller by factor 10 relative to mass densities of dark matter. But we might consider the following possible deviations from the above measurements. 1) Pulsars with excessive dispersion were excluded from the above calculations [14] so that the total electrons average density might be larger. 2) The electrons densities derived in the above measurements vary in a large range of values around their average values. The transition from partially ionized plasma to completely ionized plasma occurs at electrons density of $n_e = 0.075 \, (cm^{-3})$ which corresponds to plasma mass density of $1.88 \times 10^{-22} \, (kg/m^3)$ which is smaller from the average dark matter density given by Table 1 by only factor 4. We find that the average plasma mass density depends on the choice of pulsars in such statistics.

It is interesting also to refer to two other measurements of electrons densities: 1) For the pulsar J0835-4510 the value $n_e(cm^{-3}) = 0.265$ was obtained [8] corresponding to Hydrogen plasma mass density of $5.7 \times 10^{-22} \, (kg/m^3)$ which is approximately equal to the dark matter mass density. 2) For the pulsar JO437-4715 the value $n_e(cm^{-3}) = 0.0177$ was obtained [25] corresponding to Hydrogen plasma mass density of $4.44 \times 10^{-23} \left( kg/m^3 \right)$ (smaller from the above average dark matter mass density approximately by factor 17).

## 4. Summary and conclusion

The extensive publications on DM measurements can be used for getting average electrons densities for the plasmas in different locations of our galaxy. We find close relations between the plasma temperature and its density which follow from the heating mechanism of interstellar gas by cosmic rays. In the range or temperatures $10-10000(K)$ with corresponding densities in an inverse range of densities $10^6 - 0.2 \ (cm^{-3})$ the plasma is partially ionized, and enormous amount of spectroscopic data are available as analyzed by the ISM model. According to the present analysis there is a phase transition from partially ionized Hydrogen plasma to completely ionized plasma at temperature given approximately by $20000(K)$ as verified by solutions of Saha's equation. For temperatures above this phase transition the density of the Hydrogen plasma is inversely proportional to its temperature as obtained from cosmic rays heating mechanism. We find that such plasmas include the warm ionized medium (WIM) described as transparent plasma and for very high temperatures (around and above $10^6 (K)$) where X-rays are emitted, we have the medium as HIM. Since the WIM plasma is transparent, we found it of interest to compare the results of the DM measurement in this region with dark matter densities. Using measurements of average electrons densities obtained by DM measurements in the WIM region we found that the average mass of this transparent plasma is smaller from that of dark matter mass densities given in Table 1 by nearly one order of magnitude. We discussed certain possibilities that can reduce this difference.

**Disclosures and declarations**

**Conflict of interest:** The author has no conflicts of interest

**Author Contributions:** Investigation of Y. Ben-Aryeh

**DATA AVAILABILITY:** The data that support the finding of this study are available within the article.

**Funding:** The present study was supported by Technion-Mossad under grant No. 2007156